\documentclass[aps, prl, twocolumn, superscriptaddress,preprintnumbers,nofootinbib]{revtex4}

\usepackage{graphicx}
\usepackage{dcolumn}
\usepackage{amssymb}

\newcommand{\ben}{\begin{equation}}
\newcommand{\een}{\end{equation}}
\newcommand{\bea}{\begin{eqnarray}}
\newcommand{\eea}{\end{eqnarray}}
\newcommand{\ba}{\begin{array}}
\newcommand{\ea}{\end{array}}
\newcommand{\bit}{\begin{itemize}}
\newcommand{\eit}{\end{itemize}}

\begin{document}

\newcommand{\fd}{f_{10}}
\newcommand{\As}{A_{\mathrm{s}}}
\newcommand{\At}{A_{\mathrm{t}}}
\newcommand{\ns}{n_{\mathrm{s}}}
\newcommand{\nt}{n_{\mathrm{t}}}
\newcommand{\Obhh}{\Omega_{\mathrm{b}}h^{2}}
\newcommand{\Omhh}{\Omega_{\mathrm{m}}h^{2}}
\newcommand{\Ol}{\Omega_{\Lambda}}

\newcommand{\CMB}{\textsc{CMB }}
\newcommand{\CMBEASY}{\textsc{CMBeasy }}
\newcommand{\TT}{\textsc{tt }}
\newcommand{\TE}{\textsc{te }}
\newcommand{\EE}{\textsc{ee }}
\newcommand{\BB}{\textsc{bb }}
\newcommand{\WMAP}{\textsc{WMAP }}
\newcommand{\UETC}{\textsc{uetc }}
\newcommand{\UETCs}{\textsc{uetc}s}
\newcommand{\ETC}{\textsc{etc }}
\newcommand{\ETCs}{\textsc{etc}s }
\newcommand{\HK}{\textsc{HKP }}
\newcommand{\BBN}{\textsc{BBN }}
\newcommand{\mcmc}{\textsc{MCMC }}

\newcommand{\half}{\frac{1}{2}}

\newcommand{\clover}{\textsc{C}$\ell$\textsc{over}}

\renewcommand{\d}{{\partial}}

\newcommand{\BHKU}{BHKU}
\newcommand{\Ob}{\ensuremath{\Omega_{\mathrm b}}}
\newcommand{\Oc}{\ensuremath{\Omega_{\mathrm c}}}
\newcommand{\Omm}{\ensuremath{\Omega_{\mathrm m}}}
\newcommand{\Ochh}{\ensuremath{\Oc h^{2}}}
\newcommand{\optdepth}{\tau}
\newcommand{\Asz}{A_\mathrm{SZ}}
\newcommand{\Ap}{A_\mathrm{p}}
\newcommand{\Cl}{\mathcal{C}_{\ell}}
\newcommand{\Dl}{\mathcal{D}_{\ell}}
\newcommand{\VEV}{\PHI_{0}}
\newcommand{\PHI}{\phi}
\newcommand{\conj}{^*}
\newcommand{\FT}[1]{\tilde{#1}}
\newcommand{\vect}[1]{\mathbf{#1}}
\newcommand{\diff}{\mathrm{d}}
\newcommand{\Sr}{^{\mathrm{S}}}
\newcommand{\Vr}[1]{\!\!\stackrel{\scriptstyle{\mathrm{V}}}{_{\!\!#1}}}
\newcommand{\Tr}[1]{\!\!\stackrel{\scriptstyle{\mathrm{T}}}{_{\!#1}}}
\newcommand{\MM}{\mathcal{M}}

\newcommand{\unit}[1]{\;\mathrm{#1}}
\newcommand{\Eq}[1]{Eq. (\ref{eqn:#1})}
\newcommand{\Eqnb}[1]{Eq. \ref{eqn:#1}}
\newcommand{\Fig}[1]{Fig. \ref{fig:#1}}
\newcommand{\Sec}[1]{Sec. \ref{sec:#1}}
\newcommand{\Table}[1]{Table \ref{tab:#1}}

\newcommand{\PLSZ}{PL$_{\rm SZ}$}
\newcommand{\HZSZ}{HZ$_{\rm SZ}$}

\newcommand{\mbh}[1]{\textbf{#1}}

\newcommand{\Planck}{{\it Planck}}
\newcommand{\LCDM}{$\Lambda$CDM}

\title{Can topological defects mimic the BICEP2 B-mode signal?} 

\newcommand{\addressSussex}{Department of Physics \&
Astronomy, University of Sussex, Brighton, BN1 9QH, United Kingdom}
\newcommand{\addressGeneva}{D\'epartement de Physique Th\'eorique \& Center for Astroparticle Physics,
Universit\'e de Gen\`eve, Quai Ernest-Ansermet 24, CH-1211 Gen\`eve 4, Switzerland}
\newcommand{\addressBilbao}{Department of Theoretical Physics, University of the Basque Country UPV/EHU,
48080 Bilbao, Spain}
\newcommand{\addressEdinburgh}{Institute for Astronomy, University of Edinburgh, Royal Observatory, Edinburgh EH9 3HJ,
United Kingdom}
\newcommand{\addressHelsinki}
{Department of Physics and Helsinki Institute of Physics, PL 64, FI-00014 University of Helsinki, Finland}
\newcommand{\addressAIMS}
{African Institute for Mathematical Sciences, 6 Melrose Road, Muizenberg, 7945, South Africa}

\author{Joanes Lizarraga}
\affiliation{\addressBilbao}

\author{Jon Urrestilla}
\affiliation{\addressBilbao}

\author{David Daverio}
\affiliation{\addressGeneva}

\author{Mark Hindmarsh}
\affiliation{\addressSussex}
\affiliation{\addressHelsinki}

\author{Martin Kunz}
\affiliation{\addressGeneva}
\affiliation{\addressAIMS}

\author{Andrew R.~Liddle}
\affiliation{\addressEdinburgh}

\date{\today}

\begin{abstract}
We show that the B-mode polarization signal detected at low multipoles by BICEP2 cannot be entirely due to topological defects.  This would be incompatible with the high-multipole B-mode polarization data and also with existing temperature anisotropy data. Adding cosmic strings to a model with tensors, we find that B-modes \emph{on their own} provide a comparable  limit on the defects to that already coming from \Planck\ satellite temperature data. We note that strings at this limit give a modest improvement to the best-fit of the B-mode data, at a somewhat lower tensor-to-scalar ratio of $r \simeq 0.15$.
\end{abstract}

\maketitle


\section{Introduction}

The detection of low-multipole B-mode polarization anisotropies by the BICEP2 project \cite{Ade:2014xna} opens a new observational window on models that generate the primordial perturbations leading to structure formation. The leading candidate to explain such a B-mode signal is primordial gravitational wave (tensor) perturbations generated by the inflationary cosmology. For a tensor-to-scalar ratio $r$ of around $0.2$, these give a good match to the spectral shape in the region $\ell \simeq 40$ -- $150$, while falling some way short of the observed signal at higher multipoles for reasons yet to be uncovered.

An alternative mechanism of generating primordial B-modes is the presence of an admixture of topological defects (see e.g.\ Refs.~\cite{VilShe94,Hindmarsh:1994re,Durrer:2001cg,Copeland:2009ga,Hindmarsh:2011qj} for reviews). Many inflation scenarios, particularly of hybrid inflation type, end with a phase transition. Defect production at such a transition is natural and plausibly a sub-dominant contributor to the total temperature anisotropy. Many papers have used recent data to impose constraints on the fraction of defects, typically obtaining limits of a few percent contribution to the large-angle temperature anisotropies \cite{Wyman:2005tu,Bevis:2007gh,Battye:2010xz,Dunkley:2010ge,Urrestilla:2011gr,Avgoustidis:2011ax,Ade:2013xla}. The tensor and defect spectra were previously compared in Refs.~\cite{Urrestilla:2008jv,Mukherjee:2010ve}.

An important question then arises: does the observed B-mode polarization confirm the existence of a primordial gravitational wave background due to inflationary dynamics in the early Universe, or could it instead  be entirely due to the presence of topological defects? In this \emph{Letter} we show that topological defects alone cannot explain the BICEP2 data points. 

\begin{figure}[t]
\resizebox{1.05\columnwidth}{!}{\includegraphics{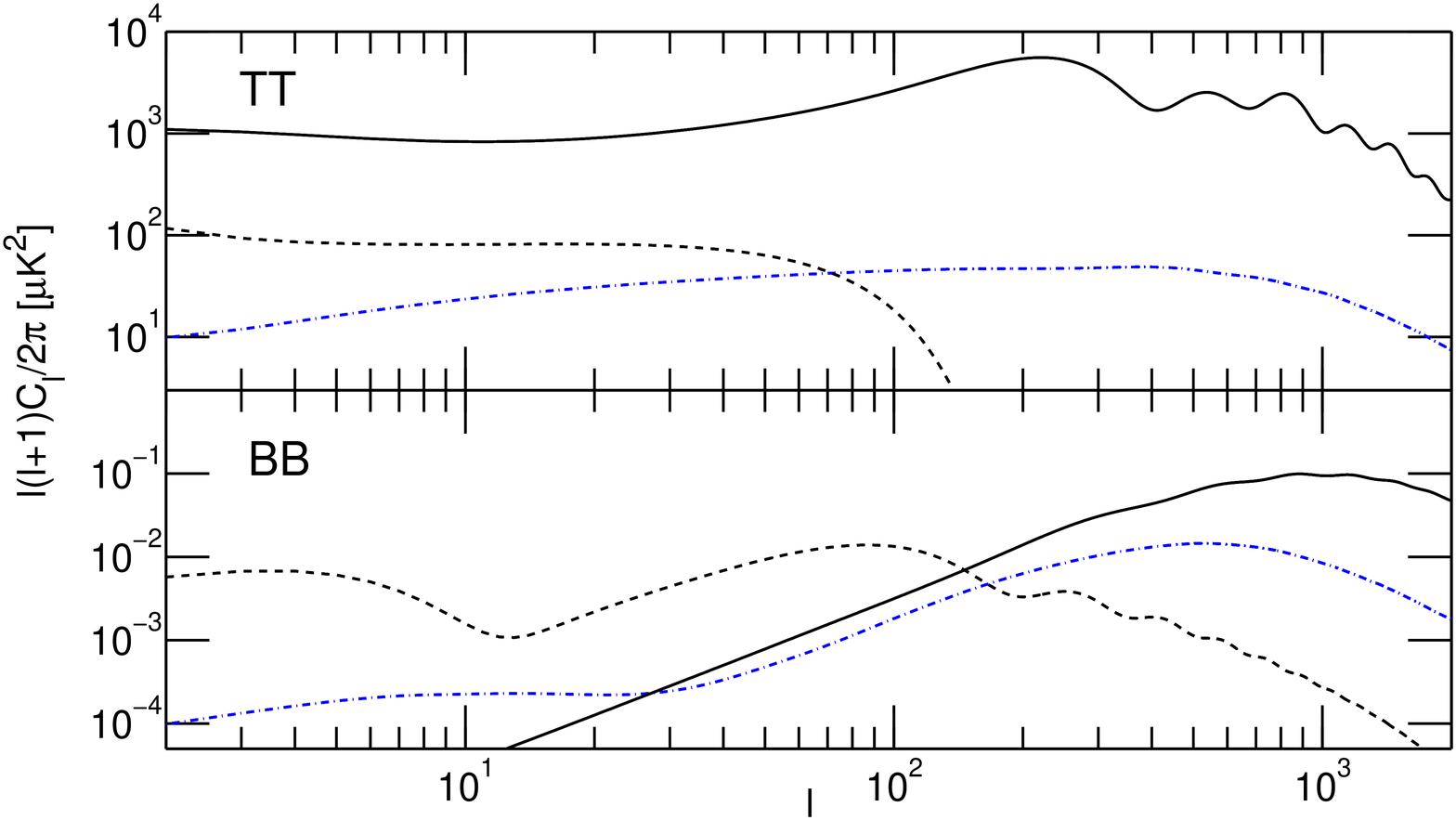}}
\caption{\label{pol}The \CMB temperature and polarization power
spectra contributions from inflationary scalar
modes (black solid), inflationary tensor modes (black dashed), and cosmic strings (blue dot-dashed) 
\cite{Bevis:2007qz}. The inflationary tensors have $r=0.2$ while the string contribution has $\fd=0.03$.}
\end{figure}

\begin{figure}[t]
\resizebox{1.05\columnwidth}{!}{\includegraphics{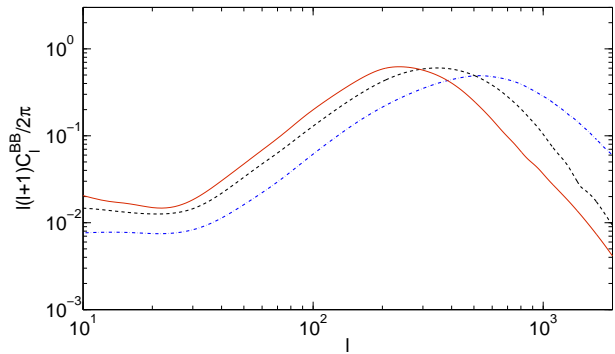}}
\caption{\label{polar} B-mode polarization power spectra for textures (solid red), semilocal strings (dashed black), and Abelian Higgs strings (dot-dash blue). All the curves are normalized to make the temperature spectra match the \Planck\ $\ell=10$ value. We see that all these types of topological defects predict similar shapes in the BICEP2 data range $30 \lesssim \ell \lesssim 300$,  though they become different for $\ell>300$. }
\end{figure}

\section{B-mode constraints from BICEP2}

As with inflationary tensors, a distinctive signature of topological defects lies in the B-mode polarization, where the signal is not masked by a dominant contribution from inflationary scalars. Figure~\ref{pol} shows a comparison of cosmic microwave background (CMB) spectra predicted from inflation with those of cosmic strings as computed via field theory simulations\footnote{Strings can also be studied in the Nambu--Goto approximation, most recently in Ref.~\cite{Blanco-Pillado:2013qja}. However, the shapes of the cosmic string CMB spectra are reasonably generic and can be understood from simple modelling \cite{Pogosian:1999np,Martins:2003vd,Battye:2010xz}. There are significant differences in other observational constraints: for a review see Ref.~\cite{Hindmarsh:2011qj}.} 
by Bevis et al.~\cite{Bevis:2007qz,Bevis:2010gj}, for a particular value of $f_{10}$ near the {\it Planck} upper limit \cite{Ade:2013xla} (where $f_{10}$ is the fractional contribution of defects to the temperature anisotropies at $\ell = 10$).  The scalar B-mode spectrum is the one inevitably produced by lensing of the scalar E-modes. In the B-mode channel the string spectrum has a quite different shape to the inflationary tensors, peaking towards smaller scales. Figure~\ref{polar} shows the B-mode polarization spectra for several classes of defects (textures, semilocal strings, and Abelian Higgs strings \cite{Urrestilla:2007sf}), showing that they share the same general shape in the multipole range of interest. We focus on cosmic strings (using the Abelian Higgs model) as a specific example for the remainder of this work.

\begin{figure}[t]
\resizebox{1.0\columnwidth}{!}{\includegraphics{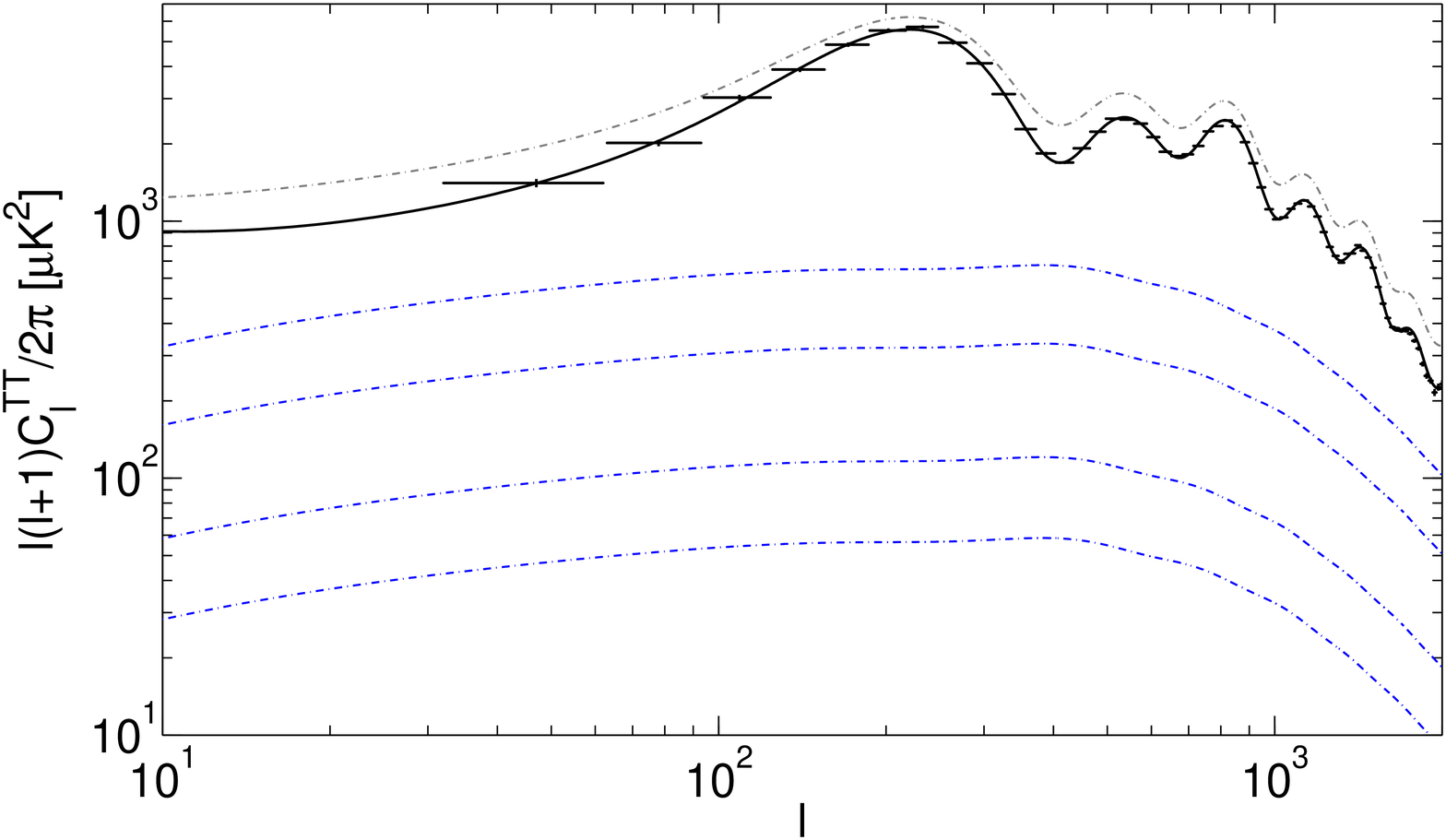}} \\
\resizebox{1.0\columnwidth}{!}{\includegraphics{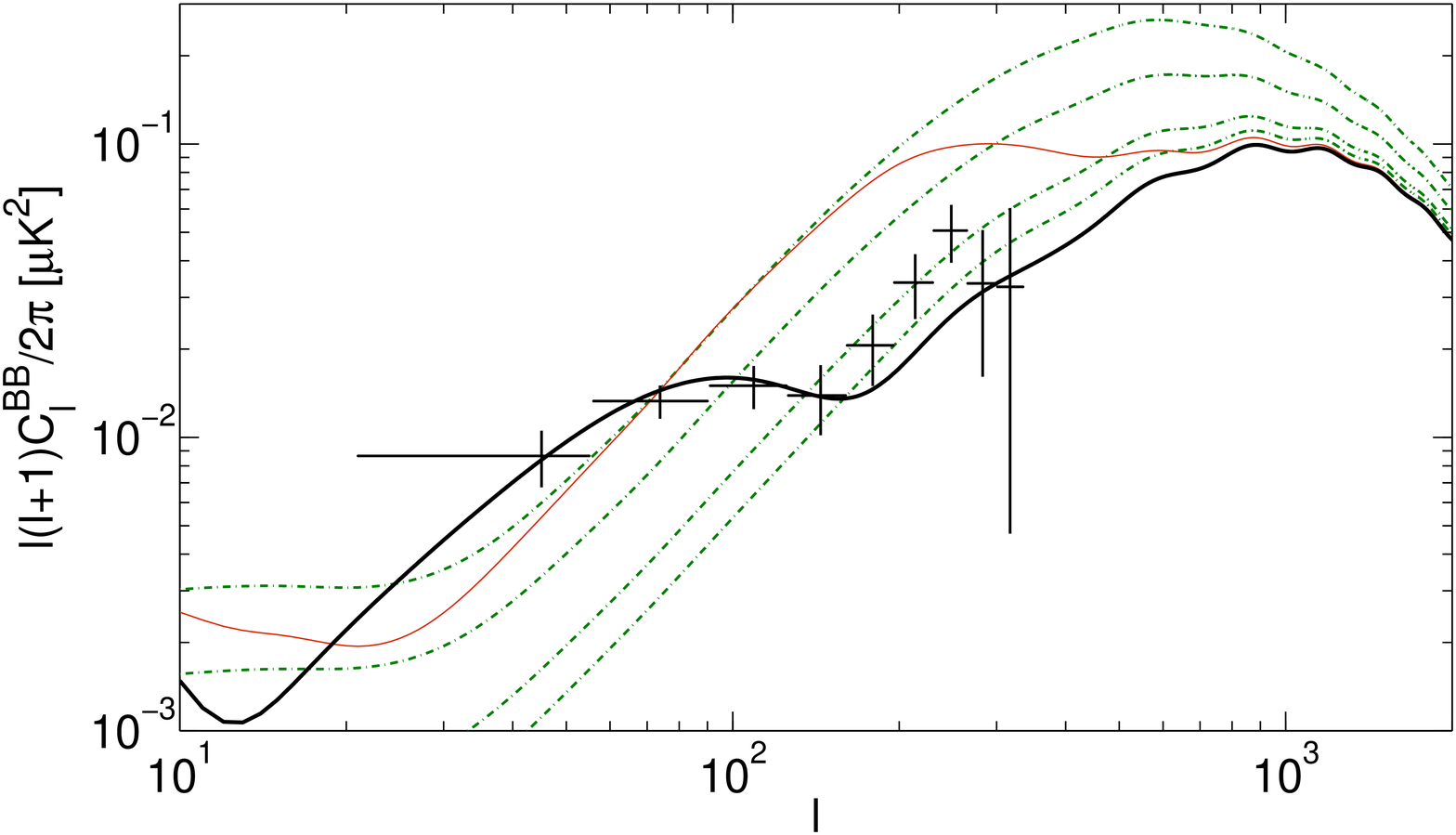}}
\caption{\label{BBr0} Temperature (upper panel) and B-mode polarization (lower panel) power spectra compared to the {\it Planck} temperature and the BICEP2 B-mode polarization data. The black curve in the upper panel is the best-fit $\Lambda$CDM model and the blue dashed lines show the contribution from strings for $f_{10} = 0.3$, $0.15$, $0.06$, and $0.03$. The green-dotted curves in the lower panel show the combined contribution from strings and the lensing of the scalar perturbations, for the same values of $f_{10}$ as in the upper panel. 
The lowest dotted curve, for $f_{10} = 0.03$, shows roughly the maximal allowed contribution from strings to the temperature power spectrum, given the \Planck\ data. 
The highest dotted curve, $f_{10} = 0.3$, matches the BICEP2 B-mode polarization at $\ell = 80$. 
The grey dashed line is the sum of the $f_{10} = 0.3$ string prediction with the \Planck\ best-fit \LCDM\ model.  The thin solid red line in the lower panel shows the combined contribution from the lensing of scalar perturbations and textures, normalized to match the $\ell=80$ BICEP2 data point. }
\end{figure}

We first attempt to match the cosmic string B-mode spectrum to the BICEP2 data, showing the result in the lower panel of Figure~\ref{BBr0}. It is clear that the  defect spectrum has the wrong shape, and could only match the low-multipole data at $\ell < 100$ by substantially over-predicting the high multipole data ($\ell >100$). In detail, we see that we need $f_{10} \simeq 0.3$ to generate the necessary power at $\ell = 80$, which in turn leads to a B-mode amplitude which is a factor of about 5 too large at higher~$\ell$. 

In addition, matching the low-multipole data requires a fractional contribution to the total TT power spectrum at $\ell=10$ far larger than the maximum allowed by \Planck\ \cite{Ade:2013xla}, as shown in the upper panel of Figure \ref{BBr0}.
We show the defect contributions to the temperature spectrum as the blue-dotted curves, with the required contributions to match the B-mode polarization amplitude at $\ell = 80$ as the highest blue-dotted curve (which corresponds to $f_{10} = 0.3$).
The solid back line is the best-fit \LCDM\ model, 
while the grey dashed line shows the sum of the $f_{10} = 0.3$ string prediction with the \Planck\ best-fit \LCDM\ model \cite{Ade:2013ktc}. 
The model in which strings match the B-mode polarization amplitude at $\ell = 80$ is clearly incompatible with the temperature data.  Allowing the parameters of the \LCDM\ model to vary does not help: 
the 95\% upper limit  from {\it Planck} is around 0.03 to 0.055 depending on the type of defect~\cite{Ade:2013xla}. 
 
We can therefore immediately conclude that defects do not provide an alternative to inflationary tensors in explaining the observed data.

\begin{figure}[t]
\resizebox{1.05\columnwidth}{!}{\includegraphics{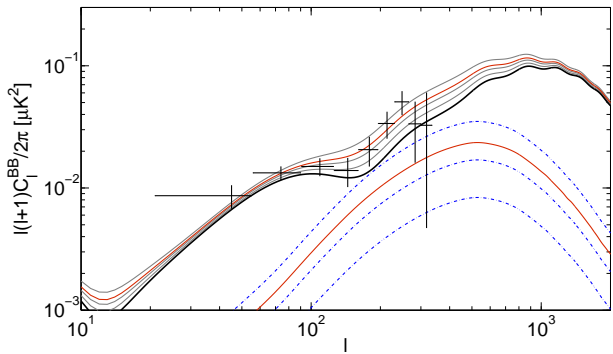}}
\caption{\label{BBr02}  A contribution from strings (blue dot-dashed) is added to the prediction from $r = 0.15$ plus scalar lensing (solid black) to give a total spectrum shown in grey. The data points are from BICEP2. From bottom to top the string fractions are 0.015, 0.03, 0.04 (highlighted in red),  and 0.06.  A marginal improvement to the overall data fit is given for a string fraction around 0.04, which is about the maximum permitted by current constraints from \Planck\ .}
\end{figure}

We can also use the B-mode data to constrain the contribution of defects to the total anisotropy in a scenario where both strings and inflationary gravitational waves contribute significantly, as anticipated in Refs.~\cite{Seljak:2006hi,Pogosian:2007gi}. In fact, because the strings contribute more substantially at higher multipoles than  inflationary tensors do, a modest admixture of defects improves the fit to the BICEP2 data; as seen in Fig.~\ref{BBr02} a string fraction of around 0.04
would explain the excess signal at $\ell \simeq 200$ (as an alternative to the more prosaic possible explanations of a foreground contribution or undiscovered systematic), while a fraction above about 0.06 is disfavoured. It is noteworthy that the first detection of the B-modes already gives a limit on defects which is competitive with that from the temperature spectrum. This conclusion can of course only strengthen if some or all of the BICEP2 signal turns out not to be cosmological.

\section{Conclusions}

If this detection of B-mode polarization is confirmed, then primordial gravitational waves appear to be a necessary addition to the standard cosmological model. However, the BICEP2 data points do not agree well with expectations at higher $\ell$. It is intriguing that an admixture of topological defects appears able to improve the fit, while reducing the tensor-to-scalar ratio to $r \simeq 0.15$. But precise quantitative statements for such a model, which would simultaneously include primordial tensors, defects, and perhaps also a running of the scalar spectral index, require a more careful numerical analysis. 

In conclusion, we have shown that topological defects alone cannot explain the BICEP2 data points, and that B-modes already give a constraint on defects competitive with that from temperature anisotropies.

\begin{acknowledgments}
JL and JU acknowledge support from the University of the Basque Country UPV/EHU (EHUA 12/11), the Basque Government (IT-559-10), the Spanish Ministry (FPA2012-34456) and the
Consolider-Ingenio  Programme CPAN (CSD2007-00042), EPI (CSD2010-00064).
DD and MK acknowledge financial support from the Swiss NSF.
MH and ARL acknowledge support 
from the Science and Technology Facilities Council (grant
numbers ST/J000477/1 and  ST/K006606/1).
\end{acknowledgments}

{\it Shortly after our article was posted on arxiv.org, a related paper \cite{Moss:2014cra}
 was posted investigating similar ideas.}

\vspace{3mm}

\bibliography{CosmicStrings.bib}

\end{document}